\documentclass{emulateapj}

\def\kms{\ifmmode{\rm km\thinspace s^{-1}}\else km\thinspace s$^{-1}$\fi}
\def\hd{HD~110555}

\journalinfo{To appear in The Astrophysical Journal, June 2007 issue}
\submitted{}

\shortauthors{Torres, Latham \& Stefanik}
\shorttitle{\hd}

\begin{document}

\title{Cross-correlation in four dimensions: Application to the
quadruple-lined spectroscopic system \hd}

\author{Guillermo Torres, David W.\ Latham and Robert P.\ Stefanik}

\affil{Harvard-Smithsonian Center for Astrophysics, 60 Garden St.,
Cambridge, MA 02138}

\email{(gtorres,dlatham,rstefanik)@cfa.harvard.edu}

\begin{abstract}

We develop a technique to measure radial velocities of stars from
spectra that present four sets of lines. The algorithm is an extension
of the two-dimensional cross-correlation method TODCOR to four
dimensions. It computes the correlation of the observed spectrum
against a combination of four templates with all possible shifts, and
allows also for the derivation of the light ratios of the
components. After testing the algorithm and demonstrating its ability
to measure Doppler shifts accurately even under conditions of heavy
line blending, we apply it to the case of the quadruple-lined system
\hd. The primary and secondary components of this previously known
visual binary ($\rho \sim 0\farcs4$) are each shown to be double-lined
spectroscopic binaries with periods of 57~days and 76~days,
respectively, making the system a hierarchical quadruple. The
secondary in the 76-day subsystem contributes only 2.5\% to the total
light, illustrating the ability of the method to measure velocities of
very faint components.

\end{abstract}

\keywords{
binaries: general --- 
binaries: spectroscopic --- 
binaries: visual ---
methods: data analysis --- 
stars: individual (\hd) ---
techniques: spectroscopic
}

\section{Introduction}
\label{sec:introduction}

The use of digital cross-correlation as an analysis tool in
spectroscopy dates back at least three decades. Early applications by
\cite{Simkin:74}, \cite{Lacy:77}, \cite{Tonry:79}, and others, showed
the power of the method for measuring accurate radial velocities, and
with many improvements and refinements over the years the technique
enjoys widespread use today \citep[for further details and a
historical perspective see, e.g.,][]{Hill:93, Kurtz:98}.  Compared to
the classical method of measuring the positions of individual spectral
lines (still occasionally used), cross-correlation offers a number of
important advantages beyond expediency, principally its ability to
make efficient use of all the information available in the spectral
window.  It is thus ideal for the analysis of low signal-to-noise
spectra, where classical methods tend to give poorer results. It is
also commonly used on composite spectra, e.g., double-lined
spectroscopic binaries, where the Doppler shifts are typically
determined from the centroids of the two correlation peaks that
correspond to the each of the components of the binary. Difficulties
arise, however, when the peaks are not well separated, and this can
lead to systematic errors in the radial velocities that bias the
amplitudes of the velocity curves and any physical quantities derived
from them (such as the masses), or it can prevent the measurement
altogether.

To address this problem \cite{Zucker:94} developed an extension of the
standard one-dimensional cross-correlation technique to two dimensions
(TODCOR), taking advantage of the properties of the Fourier
transform. TODCOR uses two templates, one for each component of the
binary, and thus the cross-correlation function (CCF) now depends on
the velocities of both stars relative to their templates.  The
location of the maximum in two-dimensional velocity space corresponds
to the Doppler shifts of the two components.  This effectively
decouples the primary from the secondary so that, if the templates are
a good match to the stars, blending between them is no longer a
concern.  A further extension of TODCOR to three dimensions was
subsequently introduced by \cite{Zucker:95} to deal with triple-lined
spectra, in which line blending is usually more common. In this case
the standard one-dimensional CCF would in principle reveal three peaks
(at favorable orbital phases), but the difficulties of measuring their
centroids by classical means can be even greater than in the
double-lined case. The success of the three-dimensional version of
TODCOR to overcome these difficulties is illustrated by a number of
applications over the last decade since its introduction
\citep{Torres:95, Jha:00, Mazeh:01, Covino:01, Torres:02, Torres:06}.

A recent study by \cite{Tokovinin:06} has shown that spectroscopic
binaries of solar-type are very often accompanied by more distant
components. The frequency of such multiple systems appears to be very
high: up to 97\% of spectroscopic binaries with periods shorter than 3
days show at least one additional companion, tapering off to 34\% for
periods longer than 12 days \citep[see Fig.~14 by][]{Tokovinin:06}.
Out of the 161 systems in their sample, 64 were found to be triple and
11 were found to be quadruple. Thus the frequency of systems with four
stars is not negligible, and it is to be expected that some fraction
of them will show lines of all four stars in the spectra, although
they may not always be easy to recognize. These will typically be
hierarchical quadruple systems composed of two relatively tight
binaries in a wide orbit around each other, with an angular separation
of order an arc second or less so that they are unresolved at the
spectrograph slit (or optical fiber entrance). The ability to measure
the radial velocities of the 4 stars in these systems is essential for
deriving properties such as their masses, as well as other orbital
characteristics, and these in turn are important for understanding the
origin and evolution of multiple systems in general, and higher
multiplicity systems in particular.

The extensive spectroscopic surveys at the Harvard-Smithsonian Center
for Astrophysics (CfA) with the CfA Digital Speedometers
\citep{Latham:92} have revealed an appreciable number of triple-lined
systems, and indeed even some cases with four sets of lines that have
until now remained unsolved because of the complexity of the analysis
and our concerns about blending. This provides the motivation for the
present work, in which we develop an extension of TODCOR to
\emph{four} dimensions, allowing the treatment of such cases. We
demonstrate the effectiveness of the method with numerical
simulations, showing that it is possible to measure the velocities of
the four stars even in situations with severe line blending that would
be hopeless using standard one-dimensional cross-correlation
techniques. We then apply the method to the real case of \hd, which
enabled us to detect for the first time the very faint 4th component
and to measure its Doppler shift, in addition to obtaining the
relative brightnesses of the stars.

\section{Description of the method}
\label{sec:description}

TODCOR as introduced by \cite{Zucker:94} is a two-dimensional
cross-correlat\-ion scheme in which the observed spectra, $f(n)$, are
cross-correlated against a composite template, $g(n)$, that is the sum
of two separate templates selected to match the properties of each of
the binary components. The template $g$ for this two-dimensional case
is given by $g = g_1(n-s_1) + \alpha\ g_2(n-s_2)$, where $s_1$ and
$s_2$ represent the shifts of the individual templates ($g_1$ and
$g_2$) needed to match the true Doppler shift of the respective
component. The coefficient $\alpha$ is the light ratio between the
secondary and the primary, which is assumed initially to be known. The
cross-correlation between $f$ and $g$ is performed for all possible
shifts of the two templates, and the resulting function of $s_1$ and
$s_2$ will typically have a global maximum, the location of which
provides the Doppler shifts of both stars. Brute-force computation of
these cross-correlations over all possible shifts is
impractical. However, \cite{Zucker:94} showed that the two-dimensional
correlation function $\mathcal{R}^{(2)} = \mathcal{R}^{(2)}(s_1, s_2)$
can be expressed analytically in terms of three \emph{one}-dimensional
correlation functions which are much more economical to compute. As
they pointed out, computing one-dimensional CCFs is an
$\mathcal{O}(N^2)$ process, where $N$ is the number of pixels in $f$,
but by using the properties of Fourier transforms the problem can be
reduced to an $\mathcal{O}(N \log N)$ process. This reduction in the
number of operations required is preserved in $\mathcal{R}^{(2)}$, and
is what makes the method practical. In the more general case in which
the light ratio is unknown, the correlation function
$\mathcal{R}^{(2)}$ also depends on $\alpha$, or $\mathcal{R}^{(2)} =
\mathcal{R}^{(2)}(s_1, s_2, \alpha)$.  An analytical expression can be
found for $\alpha$ that maximizes the correlation for each pair of
shifts $s_1$ and $s_2$, and this expression depends on the same three
one-dimensional CCFs \citep[see][]{Zucker:94}.

The extension of TODCOR to three dimensions presented by
\cite{Zucker:95} is fairly straightforward, the algebra being somewhat
more involved. The Fourier transform properties once again reduce the
computational problem to an $\mathcal{O}(N \log N)$ process.
Conceptually it is just as straightforward to extend the scheme to
four dimensions for the analysis of quadruple-lined spectroscopic
systems, although the algebra in this case is considerably more
complex. The template is now assumed to be a combination of four
separate (i.e., possibly different) templates, each with its own
shift, and three light ratios:
\begin{displaymath}
g = g_1(n-s_1) + \alpha\ g_2(n-s_2) + \beta\ g_3(n-s_3) +
 \gamma\ g_4(n-s_4)~,
\end{displaymath}
where $\beta$ and $\gamma$ are the light ratio of the tertiary and
quaternary components relative to the primary. The four-dimensional
CCF for the case in which the light ratios are known,
$\mathcal{R}^{(4)} = \mathcal{R}^{(4)}(s_1, s_2, s_3, s_4)$, can again
be expressed analytically in terms of one-dimensional correlation
functions and three light ratios, as described in the Appendix. These
one-dimensional correlations are between the observed spectrum and
each of the four templates, as well as between pairs of templates (in
all combinations).  To determine the velocities of the four stars one
simply searches for the maximum of $\mathcal{R}^{(4)}$ in the
four-dimensional space of $\{s_1, s_2, s_3, s_4\}$.  For the more
likely case in which the light ratios are unknown to begin with,
analytical expressions can be found for $\alpha$, $\beta$, and
$\gamma$ in terms of the same one-dimensional CCFs mentioned above
(see Appendix). These values can then be inserted into the expression
for $\mathcal{R}^{(4)}$. Thus the method is completely general in that
it permits in principle the determination of all unknown quantities
directly from the observed spectra. The only requirement is that the
four templates chosen for the cross-correlations be a reasonably good
match to each of the components of the quadruple system.

\section{Testing the algorithm}
\label{sec:simulations}

In order to test the method we generated synthetic composite spectra
to simulate observations of a quadruple-lined system by adding
together four calculated spectra with various relative shifts and
intensity ratios. Those same four calculated spectra were in turn used
as individual templates ($g_1, g_2, g_3, g_4$) in our four-dimensional
extension of TODCOR. All synthetic spectra were taken from a large
library of calculated spectra created by Jon Morse \citep[see
also][]{Nordstrom:94, Latham:02}, based on model atmospheres by R.\
L.\ Kurucz\footnote{Available at {\tt
http://cfaku5.cfa.harvard.edu}.}, which we also use in the next
section to analyze a real case.  These spectra cover a wavelength
range of approximately 80~\AA\ centered at 5187~\AA, and have a
resolving power of $\lambda/\Delta\lambda \approx 35,\!000$.  They
were intended for use with spectroscopic observations obtained with
the CfA Digital Speedometers, which have a narrower wavelength
coverage of only 45~\AA, so for comparison purposes we have considered
only this reduced spectral window in our tests.  Furthermore, in order
to make the simulations more realistic we have added Poisson noise
corresponding to different signal-to-noise ratios (SNR).

\begin{figure}
\epsscale{0.9}
\vskip 0.5in
\plotone{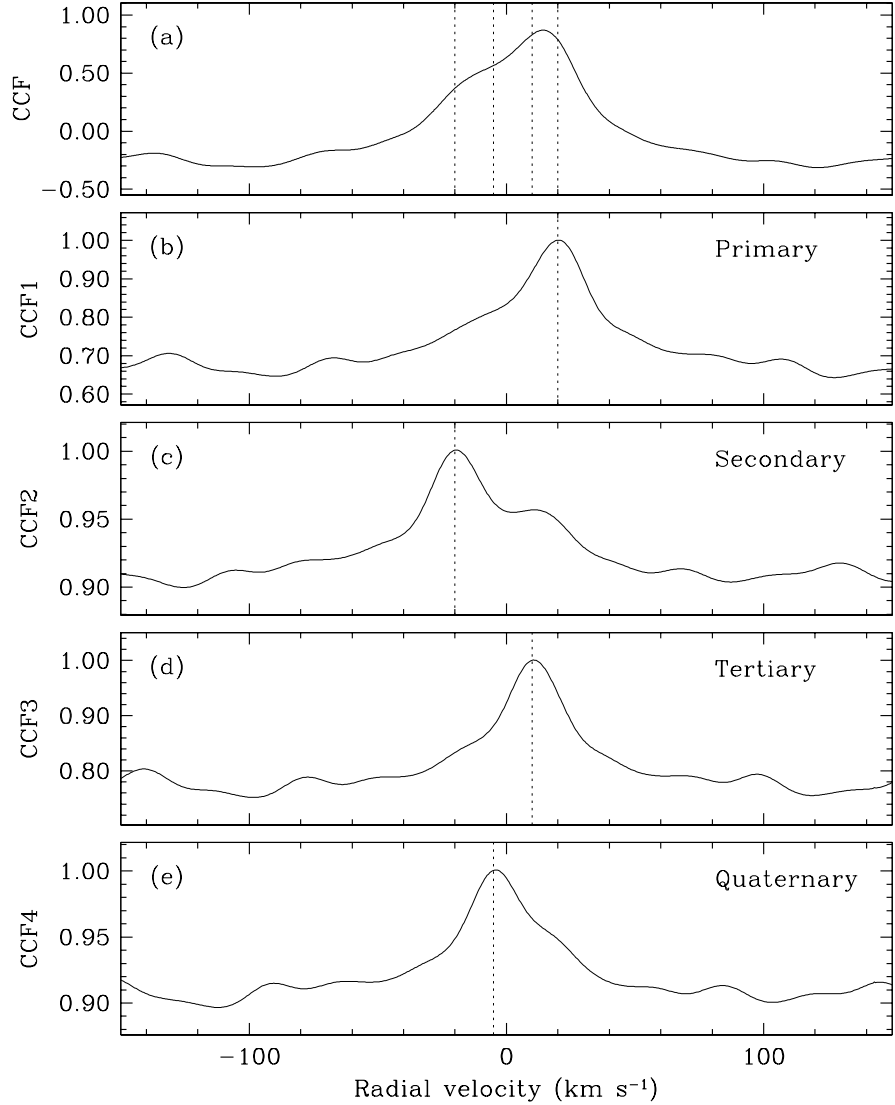}
\vskip -0.4in 

 \figcaption[]{Test of the algorithm on an artificial quadruple-lined
spectrum with strongly blended lines. ($a$) One-dimensional CCF of a
synthetic composite spectrum (G0V + G2V + G0V + G2V) against the
template corresponding to the primary star (see text). The shifts
imposed on the primary, secondary, tertiary, and quaternary are
indicated with the vertical dotted lines, and are not measurable with
standard techniques due to severe blending of the correlation
peaks. ($b$) Cross-section of the four-dimensional CCF taken at its
maximum, as a function of the primary velocity, with the velocities of
the other three components held fixed at the values that maximize the
correlation. ($c$)--($e$) Same as above, for the other three
components. \label{fig:synth1}}
\end{figure}

As a first example among the many experiments we carried out, we
consider a case with the relative shifts of the four individual stars
selected to demonstrate the ability of the algorithm to measure
velocities under conditions of significant line blending. Our
simulated quadruple system in this case is composed of non-rotating
stars of spectral types G0V, G2V, G0V, and G2V for the primary,
secondary, tertiary, and quaternary, respectively, corresponding to
effective temperatures ($T_{\rm eff}$) for the templates of 6000~K,
5750~K, 6000~K, and 5750~K. The templates were added together
(initially with no noise) with Doppler shifts of $+20~\kms$,
$-20~\kms$, $+10~\kms$, and $-5~\kms$, respectively, and with
intensity ratios of $\alpha = 0.50$ (secondary/primary), $\beta =
0.80$ (tertiary/primary), and $\gamma = 0.50$ (quaternary/primary).
Although some signs of the composite nature of this artificial
spectrum can be seen by careful visual comparison with the primary
template, it is not possible to identify the lines of the four
components, and this would only worsen in the presence of noise.
%
%
The top panel of Figure~\ref{fig:synth1} shows the standard
one-dimensional CCF we obtain by correlating the above noiseless
synthetic composite spectrum against the primary template. This and
all other one-dimensional CCFs in this paper are computed using the
IRAF\footnote{IRAF is distributed by the National Optical Astronomy
Observatories, which is operated by the Association of Universities
for Research in Astronomy, Inc., under contract with the National
Science Foundation.} task {\tt XCSAO} \citep{Kurtz:98}.  The input
velocities of each component are indicated in the figure with vertical
dotted lines. The blending of the correlation peaks is such that the
velocities of the individual stars are not measurable in this case
with standard techniques. Application of the four-dimensional
cross-correlation technique, on the other hand, is able to recover the
four velocities as well as the light ratios reliably. To show this
quantitatively and to provide at the same time a realistic measure of
the uncertainty due to noise, we have added Poisson noise to the input
composite spectrum corresponding to a SNR = 25, and generated 50
quadruple-lined spectra with different noise but with the same
shifts. We then processed these spectra as described above. The
results are shown in Table~\ref{tab:results1} (middle section under
``Test 1''), where we list the mean velocities obtained for each star
from the 50 spectra, as well as the mean light ratios. The
uncertainties attached to these means are simply the scatter of the 50
values, so they represent the typical error for a determination from a
single spectrum. Both the input velocities and the input light ratios
are recovered to well within their errors, showing the power of the
technique in this very demanding case. Cross-sections of the
four-dimensional CCF for the case of no noise are shown in the lower
panels of Figure~\ref{fig:synth1}. For example, the second panel shows
a cut along the velocity axis corresponding to the primary star, with
the velocities of the other three components held fixed at the values
that maximize the correlation function. A peak is clearly seen
precisely at the location of the input value for the velocity of that
star (dotted line). Similarly, cross-sections for each of the other
components show a prominent peak at the correct velocity.

As a second test we reduced the SNR to 10 and repeated the velocity
and light ratio determinations for the 50 simulated spectra (see
Table~\ref{tab:results1}, ``Test 2''). In this case the velocities are
typically still recovered quite reliably (within a few \kms), and
although the mean values of $\alpha$, $\beta$, and $\gamma$ are also
not far from their input values, their scatter is much
larger. However, as stated above this scatter represents the typical
error for a single measurement. In a real application there will in
general be multiple spectra of the object, so if the light ratios are
constant one can hope for a $\sqrt{N}$ gain in taking the averages.

A third experiment was carried out in which we examined the ability of
the algorithm to recover faint components in quadruple-lined
spectra. For this test we adopted templates corresponding to spectral
types G0V, G2V, G2V, and K4V (temperatures of 6000~K, 5750~K, 5750~K,
and 4750~K), and we relaxed the line blending by adopting somewhat
larger Doppler shifts of $+50~\kms$, $-50~\kms$, $+20~\kms$, and
$-20~\kms$. The SNR was set to 50 in this case, and we chose $\alpha =
0.70$ and $\beta = 0.60$. For $\gamma$ we tested a range of values
from 0.20 to 0.01, and we found that the correct velocities were
always recovered reliably by the algorithm for the primary, secondary,
and tertiary components, and that the input values of $\alpha$ and
$\beta$ were recovered as well. The velocities for the fourth star
deteriorated with decreasing $\gamma$, as expected, but were still
measurable down to $\gamma = 0.03$ ($s_4 = -22.4 \pm 1.9$~\kms\ for
the mean of 50 trials). At this light ratio the recovered value of
$\gamma$ was $0.039 \pm 0.004$, which is slightly overestimated. At a
lower light ratio of 2\% the fourth star was only detectable in about
half of these artificial spectra.

The above tests and many others we carried out show that the algorithm
performs well under conditions of severe line blending allowing the
reliable determination of radial velocities, and also demonstrate its
ability to detect faint components in quadruple-lined spectra. It is
clear that the sensitivity to faint companions depends strongly on the
degree of line blending of the component in question, as well as on
the SNR of the spectra.  In the next section we apply the method to a
real case. The performance of the algorithm will usually depend also
on how closely the four templates represent the individual stars. In
the experiments in this section the templates we have used are, by
construction, a perfect match to the components. The effects of
mismatch have not been addressed here because they depend on the
properties of the particular system under investigation, and it is
difficult to make general statements that are useful in a different
case. We discuss this further in \S\ref{sec:discussion}.
	
\section{Application to \hd}
\label{sec:application}

The relatively bright star \hd\ (also known as BD+06~2647, HIP~62034,
and ADS~8639\thinspace AB, $\alpha = 12^{\rm h} 42^{\rm m} 55\fs36$,
$\delta = +05\arcdeg 15\arcmin 59\farcs1$, J2000, spectral type G5, $V
= 8.36$) is a close ($\rho \sim 0\farcs4$) visual binary discovered by
R.\ G.\ Aitken at the Lick Observatory in 1907 \citep{Aitken:08}. It
was observed spectroscopically at the CfA as part of a large sample of
G dwarfs that have been monitored for more than 15 years. In the
course of that project we obtained a total of 48 usable spectra of
\hd\ between 1993 January and 2004 March, mostly with an echelle
spectrograph mounted on the 1.5m Wyeth reflector at the Oak Ridge
Observatory (Harvard, Massachusetts). A single echelle order spanning
45~\AA\ was recorded with a photon-counting Reticon diode array at a
mean wavelength of 5188.5~\AA, which includes the \ion{Mg}{1}~b
triplet. The SNRs of these observations range from about 11 to 42 per
resolution element of 8.5~\kms. Two of the spectra were obtained with
a nearly identical setup using the 1.5m Tillinghast reflector at the
F.\ L.\ Whipple Observatory (Mount Hopkins, Arizona). Because of the
close separation of the binary and the 1\arcsec\ width of the
spectrograph slit, the system was observed as a single object.  The
zero-point of the velocity scale was monitored by means of exposures
of the dusk and dawn sky, and small run-to-run corrections were
applied to the velocities derived below in the manner described by
\cite{Stefanik:99}.

\begin{figure}
\epsscale{1.0}
\vskip 0.3in
\plotone{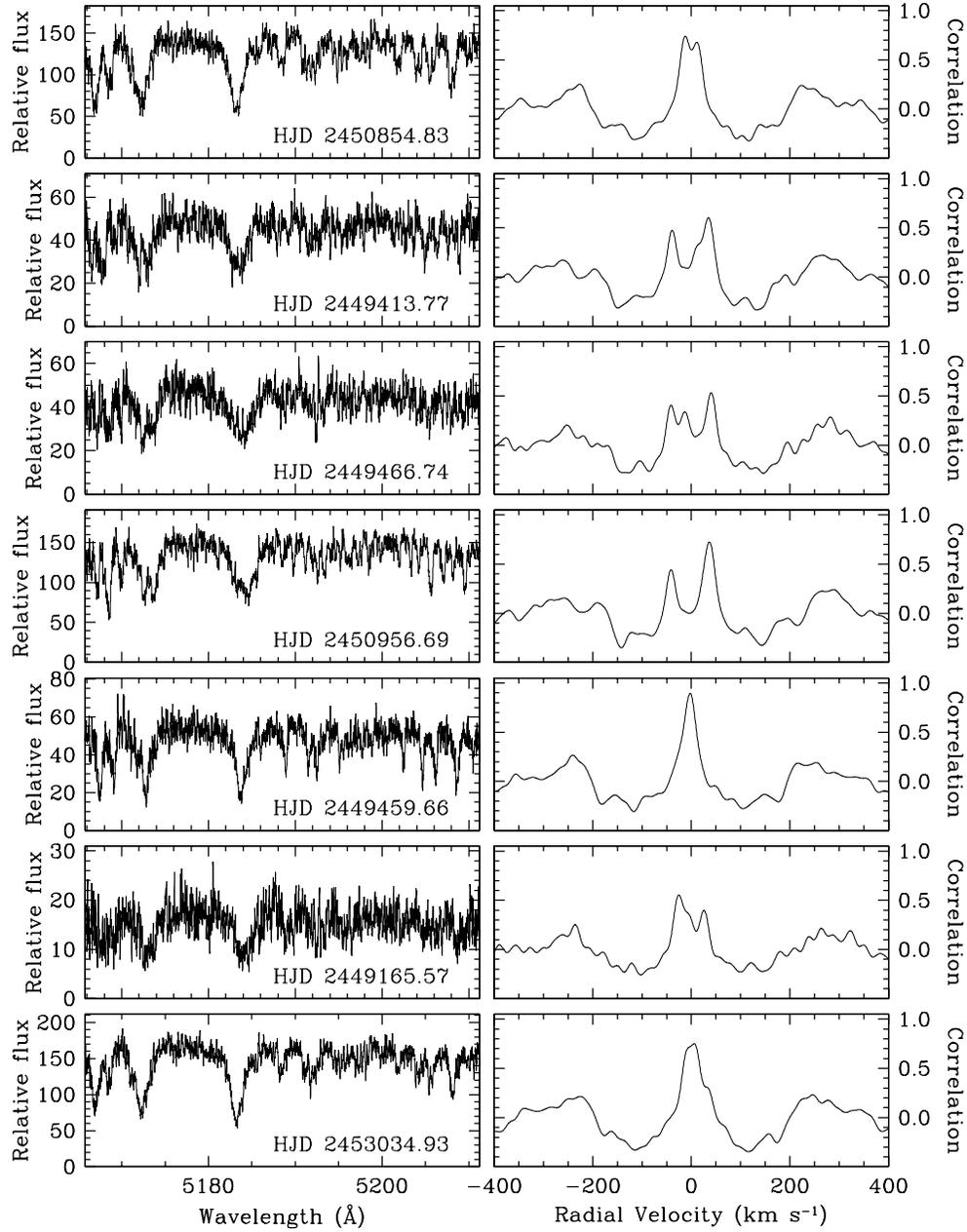}
\vskip -0.5in 

 \figcaption[]{Subset of the observations of \hd\ showing a few of our
 spectra (left) and the corresponding one-dimensional CCFs
 (right). Dates of observation are as labeled. Occasionally three
 peaks are seen in the correlation functions, although more commonly
 only one or two are visible.\label{fig:obscor}}
\end{figure}

The first few spectra of \hd\ showed obvious signs of at least two
sets of lines separated in velocity by as much as 80~\kms, which, in
view of the small angular separation of the visual pair, suggested
that at least one of the visual components is a spectroscopic binary
with a relatively short period.  Figure~\ref{fig:obscor} includes a
small sampling of our observations showing the spectra and
one-dimensional CCFs computed using a template corresponding to a G0
star.  Some of the observations even showed three sets of lines, and
the appearance of the correlation functions varied on a timescale of a
week or so.  Changes in velocity had not previously been measured for
\hd, and in fact we are only aware of a single velocity measurement in
the literature reported by \cite{Nordstrom:04} yielding $RV = -40.9
\pm 0.8$~\kms, which is, however, unlikely to be very meaningful in
view of the complicated nature of the system.

Preliminary velocity determinations were carried out initially using
the two-dimensional version of TODCOR, focusing on the two dominant
correlation peaks, and later with the three-dimensional version of the
algorithm to account for the third peak. For these analyses we adopted
solar-type templates for all stars ($T_{\rm eff} = 5750$~K), with no
rotational broadening since the composite spectra do not exhibit
significant line broadening at phases where the lines are all
blended\footnote{A $v \sin i$ measurement of 9~\kms\ was reported for
\hd\ by \cite{Nordstrom:04}, but as with the velocity estimate from
this source quoted earlier, its interpretation is difficult and we
consider it at best only as an upper limit.}.  The surface gravity was
set to $\log g = 4.5$, as appropriate for dwarfs, and solar
composition was assumed. The results showed that the system is
composed of a double-lined binary with a period of about 57 days, and
a single-lined binary with a period of $\sim$76 days. No sign of the
secondary of the latter system was apparent. The preliminary
center-of-mass velocities of these orbits were within a few \kms\ of
each other, indicating the physical association of the two binaries
and leading us to conclude that they most likely correspond to each of
the components of the visual binary. The \hd\ system is therefore a
hierarchical quadruple.

\begin{figure}
\epsscale{0.9}
\vskip 0.5in
\plotone{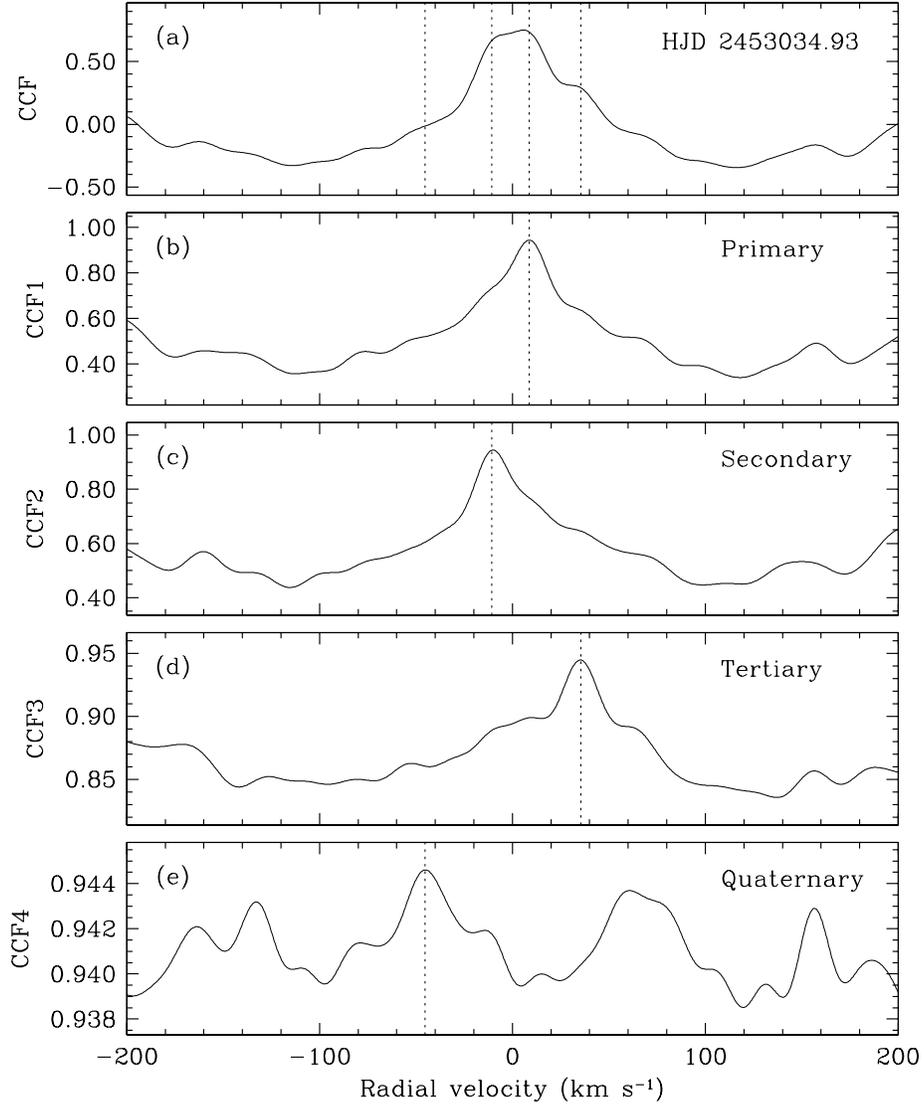}
\vskip -0.5in 

 \figcaption[]{($a$) One-dimensional cross-correlation of one of our
spectra of \hd\ (shown at the bottom of Fig.~\ref{fig:obscor}) against
the template corresponding to the primary star. The velocities
predicted from the spectroscopic orbits for the primary, secondary,
tertiary, and quaternary are indicated with the vertical dotted
lines. ($b$) Cross-section of the four-dimensional CCF taken at its
maximum, as a function of the primary velocity, with the velocities of
the other three components held fixed at the values that maximize the
correlation. ($c$)--($e$) Same as above, for the other three
components. \label{fig:realccf}}
\end{figure}

We then applied our new algorithm in an attempt to detect the fourth
star (secondary of the 76-day binary) and measure its velocity.  The
template adopted in this case was somewhat cooler ($T_{\rm eff} =
5000$~K). All 48 of our spectra showed evidence of a weak set of lines
at approximately the expected location. An example is shown in
Figure~\ref{fig:realccf}, which corresponds to the same spectrum
displayed in the bottom panel of Figure~\ref{fig:obscor}, in which the
three brighter components are quite heavily blended. The top panel of
Figure~\ref{fig:realccf} shows the one-dimensional CCF once again,
only on an expanded scale, and the lower four panels show cuts of the
four-dimensional CCF along the axes corresponding to each of the
components, as in Figure~\ref{fig:synth1}. We indicate with vertical
dotted lines the predicted velocities from our final spectroscopic
orbits, described below.  The cross-section for the quaternary is
noisy, but it does present a peak at the expected location, and the
same is true for each of our other spectra. 

\begin{figure}
\epsscale{1.0}
\vskip 0.5in
\plotone{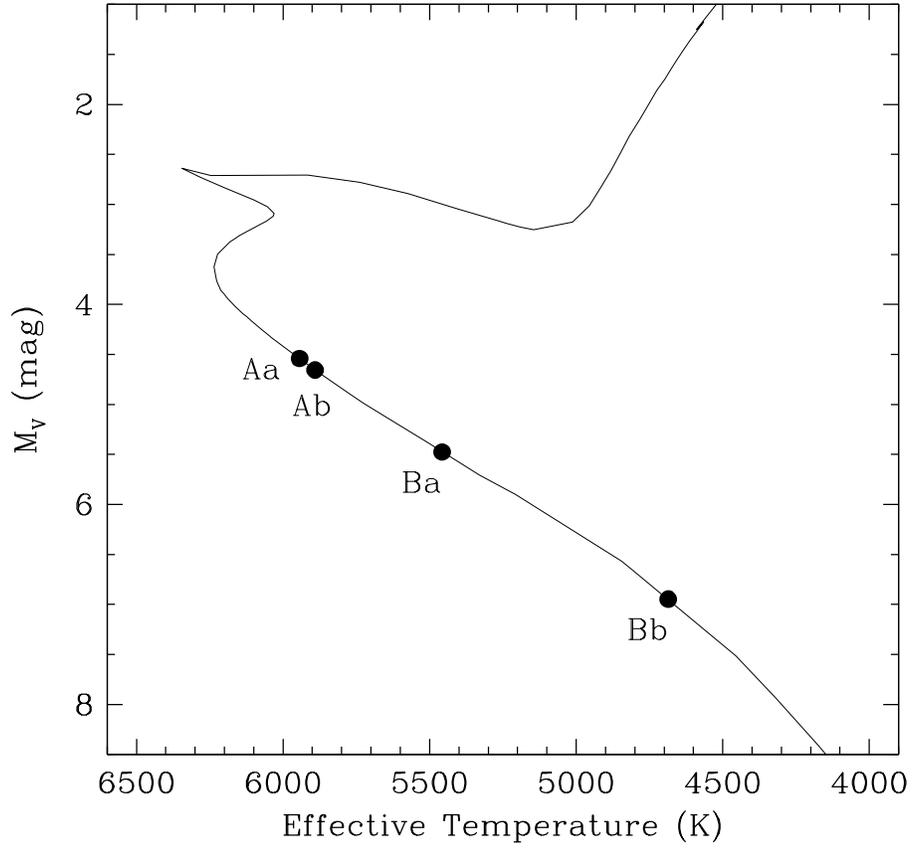}
\vskip -1.6in 

 \figcaption[]{Schematic location of the components of \hd\ in the H-R
 diagram according to our modeling, shown along with an isochrone from
 the model series by \cite{Girardi:00} for solar composition and a
 representative age of 3 Gyr. \label{fig:hr}}

\end{figure}

The preliminary light ratios we derived clearly indicated that the
57-day binary is brighter than the 76-day binary, so it must
correspond to the visual primary. We refer to this system as ``A'',
composed of stars Aa and Ab, following the usual spectroscopic
notation. Similar designations are adopted for the visual secondary,
``B''.  In order to fine-tune the templates for the four stars we made
use of model isochrones from the Padova series by \cite{Girardi:00},
combined with all available observational constraints. These include,
in addition to the three preliminary light ratios, the combined
$B\!-\!V$ color of the quadruple system as listed in the {\it
Hipparcos\/} Catalogue \citep{ESA:97}, the $V\!-\!J$, $V\!-\!H$, and
$V\!-\!K$ indices derived from the Johnson $V$ magnitude ({\it
Hipparcos\/}) and 2MASS Catalog \citep{Skrutskie:06}, properly
converted to the same photometric system as the stellar models
\citep{Carpenter:01}, the magnitude difference of the visual pair
($\Delta V$) as measured by {\it Hipparcos\/} and converted to the
visual band following \cite{Harmanec:98}, and the preliminary mass
ratios for the two spectroscopic binaries.  By requiring simultaneous
agreement with all observational quantities we were able to select
four stars from a representative model isochrone of solar metallicity
and age 3~Gyr that yield a consistent picture of the system, and allow
us to read off the effective temperatures as well as other theoretical
properties. The temperatures were close to 6000~K for Aa and Ab,
5500~K for Ba, and 4750~K for the fainter component Bb, which are the
nearest values in our library of synthetic spectra. The surface
gravities were not far from $\log g = 4.5$, as expected for dwarfs,
supporting our earlier choice of this value. The fairly close
agreement between all observational constraints and the results from
the model is shown in Table~\ref{tab:models1}, where the differences
in the last column are seen to be of order 1.3$\sigma$ or
less\footnote{The somewhat larger value of $\gamma$ predicted by the
isochrone may in fact be due at least in part to deficiencies in the
models for lower mass stars, such as missing opacities in the optical
bands \citep[see, e.g.,][]{Delfosse:00, Chabrier:05}.}.  The inferred
properties of the four stars are listed in Table~\ref{tab:models2}.
They are unevolved dwarfs, so the stellar characteristics are quite
insensitive to age for our purposes.  Their location in the H-R
diagram is illustrated in Figure~\ref{fig:hr}.  Templates with the
above parameters and zero rotational broadening were then used to
derive improved radial velocities, and the light ratios and mass
ratios also changed slightly.  One more iteration was sufficient to
reach convergence on the temperature determination for the templates,
given the relatively coarse 250~K step in our library of synthetic
spectra.  The final velocities were derived with the above template
parameters, and the light ratios obtained were $\alpha = 0.92$, $\beta
= 0.38$, and $\gamma = 0.06$, determined as described in the
Appendix. Uncertainties for these ratios are estimated to be about
0.03.

\begin{figure}
\epsscale{1.0}
\vskip 0.5in
\plotone{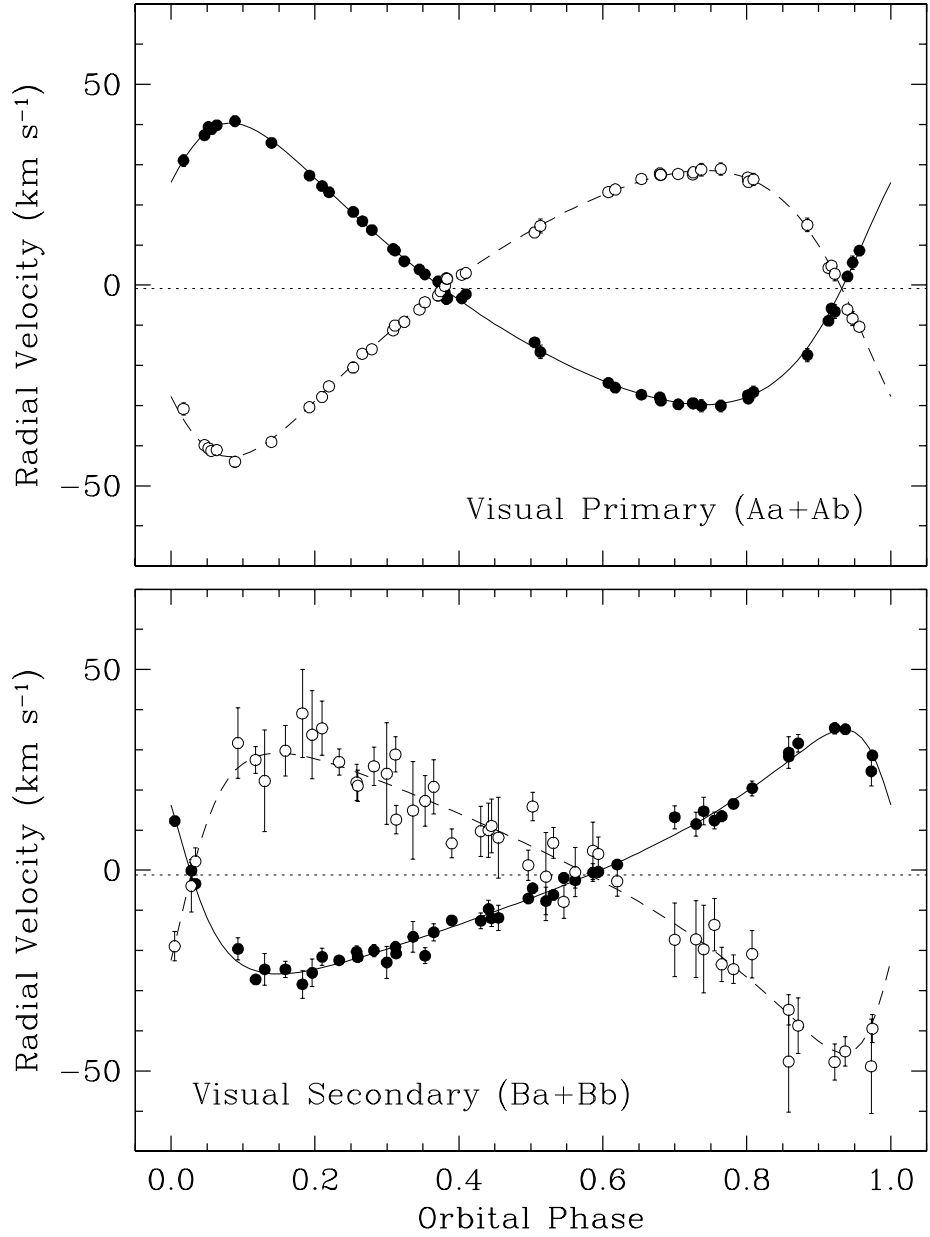}
\vskip -0.5in 

 \figcaption[]{Radial velocity measurements and computed curves for
 each of the visual components of \hd. The primary star in each
 spectroscopic binary is represented with filled circles. The
 center-of-mass velocities are indicated with dotted lines. Error bars
 for the visual primary components are typically smaller than the size
 of the points. \label{fig:orbits}}
\end{figure}

We present the velocities for the visual primary and visual secondary
components in Table~\ref{tab:rvs1} and Table~\ref{tab:rvs2}. The
uncertainties ($\sigma$) associated with these measurements reflect
both the SNR of each spectrum and the relative brightness of each
component. They were derived as a byproduct of the orbital solutions,
by assigning initial weights to the observations proportional to the
strength of each exposure, and then scaling these initial errors so
that the reduced $\chi^2$ values are near unity separately for the
primary and secondary in each subsystem.  The orbital elements we
obtain for the two spectroscopic binaries are listed in
Table~\ref{tab:elem}.  They are the period ($P$), center-of-mass
velocity ($V_0$), velocity semi-amplitudes ($K_{\rm prim}, K_{\rm
sec}$), eccentricity ($e$), longitude of periastron for the primary
($\omega_{\rm prim}$), and the time of periastron passage
($T$). Derived quantities of interest are listed as well (minimum
masses, mass ratio, and projected semimajor axes). The velocity
measurements in these fits have been weighted in the usual way
according to their uncertainties. The rms residuals of the fit for
star Ba ($\sim$1.7~\kms) and especially Bb ($\sim$5.4~\kms) are
considerably larger than those of Aa and Ab ($\sim$0.7~\kms) on
account of their faintness, which represents only about 16\% and 2.5\%
of the total brightness of the system at this wavelength. Stars Aa and
Ab contribute 42.5\% and 39\% to the total light. The velocity
measurements and corresponding orbital fits are shown graphically in
Figure~\ref{fig:orbits}.

A visual orbit determination for \hd~A and B has been published by
\cite{Ling:04} with a period of $1263 \pm 125$ yr, an angular
semimajor axis of $1\farcs915 \pm 0\farcs022$ (corresponding to
$\sim$150~AU), and a very large eccentricity of $e = 0.983 \pm 0.001$.
However, due to the small number of measurements\footnote{Nine
micrometer measurements, one speckle observation, and the measurement
from the {\it Hipparcos\/} mission. The time span of these data is
92~yr.} and their coverage of only a fraction of the orbit, it is
assigned ``Grade 5'' (``indeterminate'') in the Sixth Catalog of
Orbits of Visual Binary Stars maintained at the USNO
\citep{Hartkopf:01}, and must be regarded as very preliminary.  Based
on this solution \cite{Ling:04} reported a dynamical mass for the
system of 2.28~M$_{\sun}$, which is, however, smaller than the sum of
the minimum masses of the four stars from our spectroscopic orbits
($3.44 \pm 0.07$~M$_{\sun}$; Table~\ref{tab:elem}).  The total mass
inferred from our modeling is 3.8~M$_{\sun}$
(Table~\ref{tab:models2}).  It is quite possible that the visual
elements can in fact be improved by imposing our total mass as a
constraint. A further constraint is provided by our radial velocities
of A and B. They span only 11 yr, and we see no trend in the residuals
that would reflect motion in the outer orbit, but the difference in
the center-of-mass velocities of the two spectroscopic binary orbits,
$V_0(B) - V_0(A) = -0.26 \pm 0.23$~\kms\ (Table~\ref{tab:elem}),
provides information that is orthogonal to the astrometry, and is thus
potentially very important. We note, finally, that as a result of our
isochrone fitting described above, the inclination angles of the two
spectroscopic binaries are inferred to be $\sim$72\arcdeg\ (visual
primary) and $\sim$81\arcdeg\ (visual secondary), with uncertainties
of roughly 5\arcdeg. The inclination reported for the visual orbit is
rather similar ($85\arcdeg$).
	
\section{Discussion}
\label{sec:discussion}

The cross-correlation technique described and tested in this paper,
which is an extension of the two-dimensional algorithm TODCOR,
provides a conceptually simple and elegant way of deriving reliable
radial velocities from stellar spectra in which four sets of lines are
present. It may be thought of as a prescription for cross-correlating
the observed spectra against a composite template made of the
combination of four individual templates (one for each component) with
all possible shifts. The ability to make use of four templates
selected to match the properties of each star is one of the great
strengths of the method. Standard one-dimensional cross-correlation
methods use a single template, which may be a good representation of
one of the stars but will in general not match the others as well.
Thus the correlation will never be optimal. Another strength of our
four-dimensional algorithm is its ability to produce good results even
with low SNR spectra, as illustrated by the example of \hd, in which
the SNRs are as low as 11. Even in weak spectra such as these we were
able to measure the velocity of the faint fourth component that
contributes only 2.5\% to the light of the system (6\% of the light of
the primary), despite seeing no sign of it either in the original
spectra or in the standard one-dimensional CCFs. The method has shown
to be very effective as well in dealing with line blending, yielding
accurate radial velocities for \hd\ although the line separation
between two or more of the components is sometimes as small as a few
\kms.

Only a handful of examples of quadruple-lined spectroscopic systems
have appeared previously in the literature in which reliable radial
velocities were derived for all four stars. To our knowledge these are
XY~Leo \citep{Barden:87}, HD~221264 \citep{Willmitch:90}, ET~Boo,
VW~LMi, and TV~UMi \citep{Pribulla:06b}, AO~Vel \citep{Gonzalez:06b},
and HD~30869 \citep{Tomkin:07}, all of which contain an eclipsing pair
except for the second and last cases. A number of other spectroscopic
quadruples have been identified \citep[e.g.,][]{Pribulla:06a},
although velocity measurements have not yet been published and it is
unclear whether these systems show four sets of lines.  The material
available to the authors of the studies mentioned above was generally
of much higher quality than that employed here, with SNRs typically in
the range of 100--150. The techniques employed to derive the
velocities have varied considerably.  In the case of XY~Leo and
HD~30869 the centroids of absorption lines were measured directly, or
the observed spectra were fitted in a $\chi^2$ sense by coadding
spectra of comparison stars appropriately shifted in velocity,
rotationally broadened, and scaled in flux. In HD~221264 the lines of
the four components were separated enough that they could be isolated,
and radial velocities were derived by focusing on a set of unblended
lines of one star at a time using one-dimensional cross-correlation
methods restricted to the appropriate wavelength intervals. For
ET~Boo, VW~LMi, and TV~UMi the authors used the broadening function
(BF) formalism developed by \cite{Rucinski:02}, which reduces the
severity of blending compared to the one-dimensional cross-correlation
technique. The peaks in the BFs for these systems were fitted using
either Gaussian profiles or rotational profiles. Finally, for AO~Vel
an iterative scheme was used in which the disentangled spectra of the
components and their Doppler shifts were computed in steps
\citep[see][]{Gonzalez:06a}. All of these obviously represent useful
alternatives to the technique developed in this paper.

The advantages or disadvantages of each these methods generally depend
on the particular case under consideration, and their ability to
derive accurate radial velocities is a strong function of the degree
of line blending, the relative brightness of the components, and the
quality of the spectroscopic material, among other factors. As
demonstrated with the experiments in \S\ref{sec:simulations} the
four-dimensional extension of TODCOR presented here fares very well
even when the spectra have relatively low SNR.  Other procedures that
have been developed more specifically for disentangling the spectra of
multiple systems can also derive the Doppler shifts, such as the
method by \cite{Simon:94} that works in wavelength space, and that of
\cite{Hadrava:95} in Fourier space, although we have not yet seen an
application to quadruple-lined spectra.

An implicit requirement of the method developed in this paper is that
the four templates present a close match to the individual spectra of
the components. In our experience with the use of synthetic templates
the most important parameters are the effective temperature and the
rotational broadening.  Template mismatch is a well known issue in
digital cross-correlation that can result in biases in the radial
velocities \citep[see, e.g.,][]{Griffin:00}. Similar concerns hold in
the four-dimensional case (as in the two- and three-dimensional
versions of TODCOR). Therefore, selection of the optimal templates
must be done with care. It is possible in many cases to determine the
template parameters from the observed spectra themselves, if the SNR
is sufficient. Illustrations of this for the two- and
three-dimensional versions are given by \cite{Torres:02} and
\cite{Torres:06}. The present example of \hd\ does not lend itself to
these determinations because of the faint fourth component, which is
why we resorted to the use of models. Another possibility is to
combine our method with one of the disentangling techniques, which
should be able to produce separate spectra for each of the components
that will have higher SNRs than any of the individual spectra. These
could then be used as templates for the four-dimensional
cross-correlation scheme in order to derive the velocities. While this
would seem to be a rather obvious and elegant approach, it requires
testing to determine the sensitivity of the velocities to the details
of the disentangling.

\acknowledgements

This work was stimulated by a system similar to \hd\ brought to our
attention by I.\ Ribas and M.\ D.\ Caballero.  We thank J.\ Caruso,
R.\ J.\ Davis, and J.\ Zajac for their help in obtaining the spectra
used here, and R.\ J.\ Davis for also maintaining the CfA echelle
database. We are also grateful to T.\ Mazeh for helpful comments on an
earlier draft of this paper, and to the anonymous referee for pointing
out additional examples of quadruple-lined systems we had missed. This
work was partially supported by NSF grant AST-0406183 and NASA's
MASSIF SIM Key Project (BLF57-04). The research has made use of the
Washington Double Star Catalog maintained at the U.S.\ Naval
Observatory, of the SIMBAD database operated at CDS, Strasbourg,
France, of NASA's Astrophysics Data System Abstract Service, and of
data products from the Two Micron All Sky Survey, which is a joint
project of the University of Massachusetts and the Infrared Processing
and Analysis Center/California Institute of Technology, funded by NASA
and the NSF.

\appendix


The extension of TODCOR to four dimensions parallels the mathematical
development in the Appendices of \cite{Zucker:94} and \cite{Zucker:95}
for the two- and three-dimensional cases. The reader is referred to
those papers for some of the details. The starting point is the
definition of the one-dimensional cross-correlation function between
the object spectrum $f(n)$ and the template $g(n)$ \citep[see,
e.g.,][]{Tonry:79},
\begin{equation}
\label{eq:ccf1}
\mathcal{R}^{(1)}(s) = {\sum f(n) g(n-s)\over N \sigma_f \sigma_g}~,
\end{equation}
in which $N$ is the number of bins in the observed spectrum,
$\sigma_f$ and $\sigma_g$ are the intensity standard deviations of the
object and template spectra given by
\begin{displaymath}
\sigma_f^2 = {1\over N} \sum f(n)^2~~,~~\sigma_g^2 =
 {1\over N} \sum g(n)^2~,
\end{displaymath}
and $s$ is the shift between the two.  As is well known, the use of
the Fast Fourier Transform (FFT) allows for a very efficient
calculation of the numerator of eq.(\ref{eq:ccf1}), which is what
makes the cross-correlation method practical. This same property is
key to the success of TODCOR and its extensions.

We now consider the template $g$ to be a linear combination of four
separate templates, one for each star, with coefficients $\alpha$,
$\beta$, and $\gamma$ being the light ratios of the secondary,
tertiary, and quaternary components relative to the primary:
\begin{displaymath}
g = g_1(n-s_1) + \alpha\ g_2(n-s_2)+\beta\ g_3(n-s_3)+\gamma\ g_4(n-s_4)~.
\end{displaymath}
Substitution in eq.(\ref{eq:ccf1}) yields
\begin{equation}
\label{eq:ccf2}
\mathcal{R}^{(4)} =  {\sum f(n)\ [g_1(n-s_1)+\alpha\ g_2(n-s_2)+
\beta\ g_3(n-s_3)+\gamma\ g_4(n-s_4)]\over N \sigma_f 
\sigma_g(s_1, s_2, s_3, s_4)}~,
\end{equation}
in which
\begin{displaymath}
\sigma_g^2(s_1, s_2, s_3, s_4) = {1\over N} \sum [g_1(n-s_1)+
\alpha\ g_2(n-s_2)+\beta\ g_3(n-s_3)+\gamma\ g_4(n-s_4)]^2~.
\end{displaymath}
The numerator of eq.(\ref{eq:ccf2}) can be computed efficiently using
FFT, since the four terms have the same form as
eq.(\ref{eq:ccf1}). The standard deviation $\sigma_g(s_1, s_2, s_3,
s_4)$ in the denominator is
\begin{equation}
\label{eq:sig1}
\sigma_g^2 = \sigma^2_{g_1}+\alpha^2\sigma^2_{g_2}+
\beta^2\sigma^2_{g_3}+\gamma^2\sigma^2_{g_4}+2\alpha\sigma_{12}+
2\beta\sigma_{13}+2\gamma\sigma_{14}+2\alpha\beta\sigma_{23}+
2\alpha\gamma\sigma_{24}+2\beta\gamma\sigma_{34}~,
\end{equation}
where we have defined, as in \cite{Zucker:95},
\begin{displaymath}
\sigma_{ij}(s_j-s_i) \equiv {1\over N} \sum g_i(n-s_i)g_j(n-s_j)~.
\end{displaymath}
The first four terms on the right-hand side of eq.(\ref{eq:sig1})
include the standard deviations of the templates. The other six terms
have again the same form as the numerator in eq.(\ref{eq:ccf1}), so
they can be computed easily using FFT. If we now adopt the following
additional definitions
%
%
\begin{eqnarray}
C_i(s_i) & \equiv & {1\over N\sigma_f\sigma_{g_i}} \sum f(n)g_i(n-s_i) \label{eq:defs1} \\
C_{ij}(s_j-s_i) & \equiv & {1\over N\sigma_{g_i}\sigma_{g_j}}
\sum g_i(n)g_j\left[n-(s_j-s_i)\right]~, \label{eq:defs2}
\end{eqnarray}
we may write
\begin{eqnarray}
\label{eq:sig2}
\sigma_g^2 & =& \sigma^2_{g_1}+\alpha^2\sigma^2_{g_2}+
\beta^2\sigma^2_{g_3}+\gamma^2\sigma^2_{g_4}+ 
2(\alpha\sigma_{g_1}\sigma_{g_2} C_{12}+
\beta\sigma_{g_1}\sigma_{g_3} C_{13}+ \nonumber\\
& & \gamma\sigma_{g_1}\sigma_{g_4} C_{14}+
\alpha\beta\sigma_{g_2}\sigma_{g_3} C_{23}+
\alpha\gamma\sigma_{g_2}\sigma_{g_4} C_{24}+
\beta\gamma\sigma_{g_3}\sigma_{g_4} C_{34})~,\nonumber
\end{eqnarray}
or
\begin{displaymath}
\sigma_g^2 =  \sigma^2_{g_1}\left[ 1+\alpha^{\prime 2}+\beta^{\prime 2}+
\gamma^{\prime 2}+
2 (\alpha^{\prime}C_{12}+
\beta^{\prime}C_{13}+
\gamma^{\prime}C_{14}+
\alpha^{\prime}\beta^{\prime}C_{23}+
\alpha^{\prime}\gamma^{\prime}C_{24}+
\beta^{\prime}\gamma^{\prime}C_{34}) \right]~,
\end{displaymath}
where $\alpha^{\prime} \equiv \alpha (\sigma_{g_2}/\sigma_{g_1})$,
$\beta^{\prime} \equiv \beta (\sigma_{g_3}/\sigma_{g_1})$, and
$\gamma^{\prime} \equiv \gamma (\sigma_{g_4}/\sigma_{g_1})$. After
some algebra the final expression for the four-dimensional CCF becomes
\begin{equation}
\label{eq:ccf3}
\mathcal{R}^{(4)}={C_1+\alpha^{\prime}C_2+\beta^{\prime}C_3+
\gamma^{\prime}C_4\over\sqrt{1+\alpha^{\prime 2}+
\beta^{\prime 2}+\gamma^{\prime 2}+
2(\alpha^{\prime}C_{12}+\beta^{\prime}C_{13}+
\gamma^{\prime}C_{14}+
\alpha^{\prime}\beta^{\prime}C_{23}+
\alpha^{\prime}\gamma^{\prime}C_{24}+
\beta^{\prime}\gamma^{\prime}C_{34})}}.
\end{equation}
Thus, for known values of the three light ratios, $\mathcal{R}^{(4)} =
\mathcal{R}^{(4)}(s_1, s_2, s_3, s_4)$ is reduced to a combination of
ten one-dimensional functions that can be easily computed
(eq.[\ref{eq:defs1}] and eq.[\ref{eq:defs2}]): four of them ($C_i$, $i
= 1, \ldots, 4$) are of the observed spectrum against each of the
templates, and the other six ($C_{ij}$, $i \neq j$) are the pairwise
correlations of the four templates against each other.

In most cases the light ratios of the components are not known a
priori, but in fact they can be of considerable interest in
themselves, as illustrated by the example described in this paper.
It is therefore important to be able to estimate them from the
observed spectra. Following \cite{Zucker:95} the light ratios may be
computed directly by seeking the maximum of the correlation function
$\mathcal{R}^{(4)}$ at each combination of shifts $\{s_1, s_2, s_3,
s_4\}$. This results in a system of three equations
($\partial\mathcal{R}^{(4)}/\partial\alpha = 0$,
$\partial\mathcal{R}^{(4)}/\partial\beta = 0$,
$\partial\mathcal{R}^{(4)}/\partial\gamma = 0$) with three unknowns
($\alpha$, $\beta$, $\gamma$) which, after considerable manipulation,
can be solved analytically. We obtain the expressions
\begin{eqnarray}
\label{eq:lightratios}
\alpha & = & \left({\sigma_{g_1}\over\sigma_{g_2}}\right)
{C_1\Delta_2+C_2\Theta_4+C_3\Theta_1+C_4\Theta_2\over 
C_1\Delta_1+C_2\Delta_2+C_3\Delta_3+C_4\Delta_4} \nonumber \\
\beta & = & \left({\sigma_{g_1}\over\sigma_{g_3}}\right)
{C_1\Delta_3+C_2\Theta_1+C_3\Theta_5+C_4\Theta_3\over 
C_1\Delta_1+C_2\Delta_2+C_3\Delta_3+C_4\Delta_4} \\
\gamma & = & \left({\sigma_{g_1}\over\sigma_{g_4}}\right)
{C_1\Delta_4+C_2\Theta_2+C_3\Theta_3+C_4\Theta_6\over 
C_1\Delta_1+C_2\Delta_2+C_3\Delta_3+C_4\Delta_4}~, \nonumber
\end{eqnarray}
where we have defined
\begin{eqnarray}
\Theta_1 & = & C_{12}C_{13}+C_{14}^2C_{23}+C_{24}C_{34}-
C_{12}C_{14}C_{34}-C_{13}C_{14}C_{24}-C_{23} \nonumber\\
\Theta_2 & = & C_{12}C_{14}+C_{13}^2C_{24}+C_{23}C_{34}-
C_{12}C_{13}C_{34}-C_{13}C_{14}C_{23}-C_{24} \nonumber\\
\Theta_3 & = & C_{13}C_{14}+C_{12}^2C_{34}+C_{23}C_{24}-
C_{12}C_{13}C_{24}-C_{12}C_{14}C_{23}-C_{34} \nonumber\\
\Theta_4 & = & 1-C_{13}^2-C_{14}^2-C_{34}^2+2C_{13}C_{14}C_{34} \nonumber\\
\Theta_5 & = & 1-C_{12}^2-C_{14}^2-C_{24}^2+2C_{12}C_{14}C_{24} \nonumber\\
\Theta_6 & = & 1-C_{12}^2-C_{13}^2-C_{23}^2+2C_{12}C_{13}C_{23}\nonumber
\end{eqnarray}
and
\begin{eqnarray}
\Delta_1 & = & 1-C_{23}^2-C_{24}^2-C_{34}^2+2C_{23}C_{24}C_{34} \nonumber\\
\Delta_2 & = & C_{12}C_{34}^2+C_{13}C_{23}+C_{14}C_{24}-
C_{13}C_{24}C_{34}-C_{14}C_{23}C_{34}-C_{12} \nonumber\\
\Delta_3 & = & C_{12}C_{23}+C_{13}C_{24}^2+C_{14}C_{34}-
C_{12}C_{24}C_{34}-C_{14}C_{23}C_{24}-C_{13} \nonumber\\
\Delta_4 & = & C_{12}C_{24}+C_{13}C_{34}+C_{14}C_{23}^2-
C_{12}C_{23}C_{34}-C_{13}C_{23}C_{24}-C_{14}~.\nonumber
\end{eqnarray}
After rescaling to convert to $\alpha^{\prime}$, $\beta^{\prime}$, and
$\gamma^{\prime}$, as indicated above, the values obtained from
eqs.(\ref{eq:lightratios}) can be substituted in eq.(\ref{eq:ccf3}) to
compute the value of the correlation for the optimal values of the
three light ratios at each set of shifts $\{s_1, s_2, s_3, s_4\}$.

\clearpage

\begin{deluxetable}{lcc}
\tablecolumns{3}
\tablewidth{0pt}
\tablecaption{Tests of the algorithm using synthetic spectra.\label{tab:results1}}
\tablehead{\colhead{} & \colhead{Radial velocity} & \colhead{Light ratio} \\
\colhead{~~~~~~~Component~~~~~~~} & \colhead{(\kms)} & \colhead{($\alpha$,
 $\beta$, $\gamma$)}}
\startdata
\noalign{\vskip -5pt}
\sidehead{Input values}
~~~Primary\dotfill    & $+20.00$\phs & \nodata \\
~~~Secondary\dotfill  & $-20.00$\phs & 0.50 \\
~~~Tertiary\dotfill   & $+10.00$\phs & 0.80 \\
~~~Quaternary\dotfill & \phn$-5.00$\phs & 0.50 \\
\sidehead{Output from Test 1: SNR = 25}
~~~Primary\dotfill    & $+20.02~\pm~0.47$\phn\phs & \nodata \\
~~~Secondary\dotfill  & $-19.67~\pm~0.38$\phn\phs & $0.49~\pm~0.06$ \\
~~~Tertiary\dotfill   & $+10.13~\pm~0.86$\phn\phs & $0.82~\pm~0.14$ \\
~~~Quaternary\dotfill & $ -4.61~\pm~0.75$\phs     & $0.50~\pm~0.06$ \\
\sidehead{Output from Test 2: SNR = 10}
~~~Primary\dotfill    & $+20.17~\pm~1.44$\phn\phs & \nodata \\
~~~Secondary\dotfill  & $-19.37~\pm~1.12$\phn\phs & $0.56~\pm~0.27$ \\
~~~Tertiary\dotfill   & \phn$+9.97~\pm~2.54$\phn\phs & $1.01~\pm~0.70$ \\
~~~Quaternary\dotfill & $ -4.69~\pm~2.24$\phs     & $0.57~\pm~0.32$ \\
\enddata

\end{deluxetable}


\begin{deluxetable}{lccc}
\tabletypesize{\footnotesize}
\tablewidth{0pc}
\tablecaption{Results of a model fit to the observational constraints
for \hd, in order to infer the effective temperatures of the four
components.\label{tab:models1}}
\tablehead{\colhead{~~~~~~~~~~~~~Parameter~~~~~~~~~~~~~} & \colhead{Observed value} &
\colhead{Model result} & \colhead{($O-C$)/$\sigma$}}
\startdata
$\alpha$\dotfill             &   $0.92~\pm~0.03$      &  0.90\tablenotemark{a}   & $+$0.67\phs  \\
$\beta$\dotfill              &   $0.38~\pm~0.03$      &  0.41\tablenotemark{a}   & $-$1.00\phs  \\
$\gamma$\dotfill             &   $0.06~\pm~0.03$      &  0.10\tablenotemark{a}   & $-$1.33\phs  \\
Combined $V$ (mag)\dotfill   &   $8.36~\pm~0.01$\tablenotemark{b}       &  8.358 & $+0.20$\phs  \\
Combined $B\!-\!V$ (mag)\dotfill &   $0.658~\pm~0.019$\tablenotemark{b} &  0.667 & $-$0.47\phs  \\
Combined $V\!-\!J$ (mag)\dotfill &   $1.157~\pm~0.029$\tablenotemark{c} &  1.160 & $-$0.10\phs  \\
Combined $V\!-\!H$ (mag)\dotfill &   $1.525~\pm~0.037$\tablenotemark{c} &  1.514 & $+0.30$\phs  \\
Combined $V\!-\!K$ (mag)\dotfill &   $1.552~\pm~0.031$\tablenotemark{c} &  1.552 & 0.00 \\
$\Delta V$ (mag)\dotfill     &   $1.33~\pm~0.04$\tablenotemark{d}      &  1.383  & $-$1.32\phs  \\
$\pi$ (mas)\dotfill          &   \phantom{\tablenotemark{b}}$12.45~\pm~1.82$\tablenotemark{b}\phn &  11.05 & $+$0.77\phs \\
\enddata 
\tablenotetext{a}{Values interpolated between the $B$ and $V$ bands to
match the mean wavelength of the observed light ratios (5188.5~\AA).}
\tablenotetext{b}{As listed in the {\it Hipparcos\/} Catalogue \citep{ESA:97}.}
\tablenotetext{c}{Derived from the Johnson $V$ magnitude and $JHK_s$
from 2MASS, and converted to the \cite{Bessell:88} photometric system
adopted for the isochrones using the transformations by
\cite{Carpenter:01}.}
\tablenotetext{d}{The original {\it Hipparcos\/} measurement of the
magnitude difference between the visual components of \hd\ in the
$H_p$ band ($\Delta H_p = 1.35 \pm 0.03$) has been transformed here to
the $V$ band using the relations by \cite{Harmanec:98}.}
\tablecomments{The model fit is based on a 3~Gyr isochrone from the
series by \cite{Girardi:00} for solar metallicity. The mass ratios for
the two spectroscopic binaries have been adopted from
Table~\ref{tab:elem} as additional constraints. As an additional
check, the parallax resulting from the models (average of estimates in
four passbands) is inferred by comparison of the predicted integrated
absolute magnitudes in $V\!JHK$ with the apparent magnitudes.}
\end{deluxetable}


\begin{deluxetable}{lcccc}
\tablewidth{0pc}
\tablecaption{Properties for the four stars in \hd\ inferred from
models.\label{tab:models2}}
\tablehead{\colhead{} & \colhead{Mass} & \colhead{$T_{\rm eff}$} &
           \colhead{$\log g$} & \colhead{$M_V$} \\
           \colhead{~~~~~~~~Component~~~~~~~~} & \colhead{(M$_{\sun}$)} &
           \colhead{(K)} & \colhead{(cgs)} & \colhead{(mag)}}
\startdata
Primary (Aa)\dotfill     & 1.065 & 5944 & 4.393 & 4.540 \\
Secondary (Ab)\dotfill   & 1.048 & 5891 & 4.414 & 4.655 \\
Tertiary (Ba)\dotfill    & 0.930 & 5457 & 4.527 & 5.475 \\
Quaternary (Bb)\dotfill  & 0.758 & 4686 & 4.634 & 6.947 \\
\enddata 
\tablecomments{Results are based on a comparison of the observed
magnitudes and colors of \hd\ with a model isochrone from
\cite{Girardi:00}, for solar composition and a representative age of
3~Gyr (see text and Table~\ref{tab:models1}).}
\end{deluxetable}

\clearpage

\begin{deluxetable}{lccccccc}
\tabletypesize{\footnotesize}
\tablewidth{0pt}
\tablecaption{Radial velocity measurements for the visual primary (A =
Aa + Ab) of \hd, in the heliocentric frame.\label{tab:rvs1}}
\tablehead{\colhead{HJD} & 
           \colhead{RV$_{\rm Aa}$} & \colhead{$\sigma_{\rm Aa}$} &
           \colhead{$(O-C)_{\rm Aa}$} &
           \colhead{RV$_{\rm Ab}$} & \colhead{$\sigma_{\rm Ab}$} &
           \colhead{$(O-C)_{\rm Ab}$} & \colhead{} \\
           \colhead{~~~(2,400,000+)~~~} &
           \colhead{($\kms$)} & \colhead{($\kms$)} & \colhead{($\kms$)} &
           \colhead{($\kms$)} & \colhead{($\kms$)} & \colhead{($\kms$)} &
           \colhead{Phase}
           }
\startdata
48997.8913\dotfill & $-$17.42\phs &  1.62 & $-$1.80\phs & $+$14.99\phs & 1.66 & $+$0.86\phs &  0.8844 \\
49025.7939\dotfill &  \phn$+$0.91\phs &  1.22 & $+$0.85\phs &  \phn$-$2.60\phs & 1.25 & $-$0.80\phs &  0.3711 \\
49046.7522\dotfill & $-$30.05\phs &  1.56 & $-$0.28\phs & $+$28.76\phs & 1.60 & $+$0.24\phs &  0.7368 \\
49058.8167\dotfill &  \phn$+$5.61\phs &  1.69 & $+$0.73\phs &  \phn$-$8.37\phs & 1.73 & $-$1.66\phs &  0.9472 \\
49062.8294\dotfill & $+$31.03\phs &  1.46 & $-$0.16\phs & $-$30.83\phs & 1.50 & $+$2.62\phs &  0.0172 \\
\enddata
\tablecomments{Table~\ref{tab:rvs1} is published in its entirety in
the electronic edition of the {\it Astrophysical Journal}. A portion
is shown here for guidance regarding its form and content.}
\end{deluxetable}


\begin{deluxetable}{lccccccc}
\tabletypesize{\footnotesize}
\tablewidth{0pt}
\tablecaption{Radial velocity measurements for the visual secondary (B
= Ba + Bb) of \hd, in the heliocentric frame.\label{tab:rvs2}}
\tablehead{\colhead{HJD} & 
           \colhead{RV$_{\rm Ba}$} & \colhead{$\sigma_{\rm Ba}$} &
           \colhead{$(O-C)_{\rm Ba}$} &
           \colhead{RV$_{\rm Bb}$} & \colhead{$\sigma_{\rm Bb}$} &
           \colhead{$(O-C)_{\rm Bb}$} & \colhead{} \\
           \colhead{~~~(2,400,000+)~~~} &
           \colhead{($\kms$)} & \colhead{($\kms$)} & \colhead{($\kms$)} &
           \colhead{($\kms$)} & \colhead{($\kms$)} & \colhead{($\kms$)} &
           \colhead{Phase}
           }
\startdata
48997.8913\dotfill & $-$16.59\phs & 3.81 & $+$0.88\phs & $+$14.87\phs & 12.16 &  \phn$-$4.05\phs & 0.3364 \\
49025.7939\dotfill & $+$13.21\phs & 2.86 & $+$4.37\phs & $-$17.32\phs &  9.14 &  \phn$-$3.96\phs & 0.6999 \\
49046.7522\dotfill & $+$24.64\phs & 3.67 & $-$4.86\phs & $-$48.85\phs & 11.71 & $-$10.14\phs & 0.9730 \\
49058.8167\dotfill & $-$24.71\phs & 3.97 & $+$0.86\phs & $+$22.26\phs & 12.65 &  \phn$-$6.61\phs & 0.1302 \\
49062.8294\dotfill & $-$28.42\phs & 3.43 & $-$3.15\phs & $+$39.04\phs & 10.96 & $+$10.54\phs & 0.1825 \\
\enddata
\tablecomments{Table~\ref{tab:rvs2} is published in its entirety in
the electronic edition of the {\it Astrophysical Journal}. A portion
is shown here for guidance regarding its form and content.}
\end{deluxetable}


\begin{deluxetable}{lcc}
\tablecaption{Orbital elements for the two visual components of
\hd.\label{tab:elem}}
\tablehead{\colhead{~~~~~~~~~~~~Parameter~~~~~~~~~~~~} & 
\colhead{Visual Primary (Aa+Ab)} & \colhead{Visual Secondary (Ba+Bb)}
}
\startdata
\noalign{\vskip -3pt}
\sidehead{Adjusted quantities} \\
\noalign{\vskip -9pt}
~~~~P (days)\dotfill                  &    $57.32244~\pm~0.00095$\phn &  $76.7489~\pm~0.0059$\phn \\
~~~~$V_0$ (\kms)\dotfill              &     $-0.865~\pm~0.069$\phs    &  $-1.13~\pm~0.22$\phs \\
~~~~$K_{\rm prim}$ (\kms)\dotfill     &       $35.07~\pm~0.15$\phn    &    $30.47~\pm~0.40$\phn \\
~~~~$K_{\rm sec}$ (\kms)\dotfill      &       $35.66~\pm~0.15$\phn    &    $37.39~\pm~1.17$\phn \\
~~~~$e$\dotfill                       &      $0.3030~\pm~0.0028$  &   $0.4970~\pm~0.0091$ \\
~~~~$\omega_{\rm prim}$ (deg)\dotfill &     $305.44~\pm~0.76$\phm{22}     &     $67.6~\pm~1.5$\phn \\
~~~~$T$ (HJD$-$2,400,000)\dotfill     &   $50437.58~\pm~0.11$\phm{2222}     &  $50430.31~\pm~0.20$\phm{2222} \\
\sidehead{Derived quantities} \\
\noalign{\vskip -9pt}
~~~~$M_{\rm prim} \sin^3 i$ (M$_{\sun}$)\dotfill & $0.9169~\pm~0.0089$  &     $0.895~\pm~0.059$ \\
~~~~$M_{\rm sec} \sin^3 i$ (M$_{\sun}$)\dotfill  & $0.9019~\pm~0.0087$  &     $0.729~\pm~0.030$ \\
~~~~$q \equiv M_{\rm sec}/M_{\rm prim}$\dotfill  & $0.9837~\pm~0.0058$  &     $0.815~\pm~0.027$ \\

~~~~$a_{\rm prim} \sin i$ ($10^6$ km)\dotfill    &  $26.35~\pm~0.11$\phn    &     $27.90~\pm~0.34$\phn \\
~~~~$a_{\rm sec} \sin i$ ($10^6$ km)\dotfill     &  $26.78~\pm~0.11$\phn    &     $34.24~\pm~1.06$\phn \\
~~~~$a \sin i$ (R$_{\sun}$)\dotfill               &  $76.34~\pm~0.23$\phn    &      $89.3~\pm~1.6$\phn \\
\sidehead{Other quantities pertaining to the fit} \\
\noalign{\vskip -9pt}
~~~~Time span (days)\dotfill              &             4071.8     &      4071.8 \\
~~~~Orbital cycles\dotfill                &              71.0      &       53.1 \\
~~~~$\sigma_{\rm prim}$ (\kms)\dotfill    &              0.72      &       1.68 \\
~~~~$\sigma_{\rm sec}$ (\kms)\dotfill     &              0.73      &       5.37 \\
\enddata
\end{deluxetable}

%
%


\clearpage

\clearpage

\clearpage

\clearpage

\end{document}